# SPIN TRACKING OF POLARIZED PROTONS IN THE MAIN INJECTOR AT FERMILAB*

M. Xiao[†], Fermilab, Batavia, IL 60565, USA
W. Lorenzon and C. Aldred, University of Michigan, Ann Arbor, MI 48109-1040, USA

*Abstract*

The Main Injector (MI) at Fermilab currently produces high-intensity beams of protons at energies of 120 GeV for a variety of physics experiments. Acceleration of polarized protons in the MI would provide opportunities for a rich spin physics program at Fermilab. To achieve polarized proton beams in the Fermilab accelerator complex, shown in Fig.1.1, detailed spin tracking simulations with realistic parameters based on the existing facility are required. This report presents studies at the MI using a single 4-twist Siberian snake to determine the depolarizing spin resonances for the relevant synchrotrons. Results will be presented first for a perfect MI lattice, followed by a lattice that includes the real MI imperfections, such as the measured magnet field errors and quadrupole misalignments. The tolerances of each of these factors in maintaining polarization in the Main Injector will be discussed.

## INTRODUCTION

The Main Injector is a multi-purpose synchrotron [1] which ramps up the proton beam from a kinetic energy of 8 GeV to 120 GeV. It provides neutrino beams for the MINOS, MINERvA and NOvA experiments, as well as the future Long-Baseline Neutrino Facility and Deep Underground Neutrino Experiment. It will also provide muon beams for Fermilab's Muon g-2 and Mu2e experiments. It delivers beam to the SeaQuest fixed-target experiment and to a dedicated facility for testing of detector technologies.

The acceleration of polarized protons in the MI was initially studied with the use of two superconducting helical dipole Siberian snakes. However, in 2012 it was discovered that there was no longer sufficient space in the MI to place two Siberian snakes at opposite sides of the ring [2]. A solution using one 4-twist Helical Snake in the MI [3] was found that seemed promising to provide polarized proton beams to the experiments. Spin tracking studies in the MI became necessary to reveal if it was possible or not in practice to produce and maintain a polarized proton beam in the Fermilab accelerators using single Siberian snakes in the larger synchrotrons. This report presents studies to determine the intrinsic spin resonance strengths for the relevant synchrotrons using a perfect lattice. This is followed by the implementation of various realistic imperfections, such as magnet field errors and quadrupole misalignments, into the MI lattice to study the tolerances of closed orbit corrections in maintaining polarization. All results presented here assume that the Siberian snake is a point-like spin flipper. The simulation using a single 4-twist helical dipole and its imperfection will be discussed at a later stage.

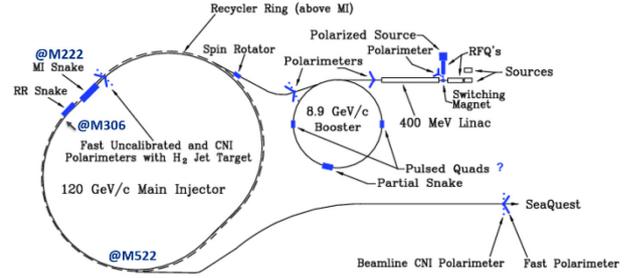

Fig. 1.1 Main Injector accelerator complex conceptual layout showing equipment needed for polarized proton beam (in blue).

## SPIN DANAMIC OF THE POLARIZED PROTON

For a beam of particles, the polarization vector is defined as the ensemble average of spin vectors. The evolution of the spin vector of a beam of polarized protons in external magnetic fields is governed by the Thomas-BMT equation [4]

$$\frac{d\vec{S}}{dt} = \frac{e}{\gamma m}\vec{S} \times \left[(1+G\gamma)\vec{B}_\perp + (1+G)\vec{B}_{//} + (G\gamma + \frac{\gamma}{\gamma+1})\frac{\vec{E}\times\vec{\beta}}{c}\right], \quad (2.1)$$

where the polarization vector $\vec{S}$ is expressed in the frame that moves with the particle. $\vec{B}_\perp$ and $\vec{B}_{//}$ are the transverse and longitudinal components of the magnetic fields in the laboratory frame with respect to the velocity $\vec{\beta}c$ of the particle. The vector $\vec{E}$ stands for the electric field, G is the anomalous gyromagnetic g-factor, and $\gamma mc^2$ is the energy of the moving particle. In a pure magnetic field, $\vec{E}=0$.

In the SU(2) representation, the spin vector can be expressed with two-component spinor $\psi = (\psi_1, \psi_2)^T$ where $\psi_1, \psi_2$ are complex numbers. The conversion between SU(2) and SO(3) is

$$\vec{S} = \psi^+\vec{\sigma}\psi \quad (2.2)$$

where $\vec{\sigma} = (\sigma_1, \sigma_2, \sigma_3)$ are Pauli matrices. Due to the unitarity of the spin vector, $P = |\psi_1|^2 + |\psi_2|^2$, P is the polarization. *P=1* for a single particle. In spinor notation, the T-BMT equation can be written as

$$\frac{d}{dt}\psi = -\frac{i}{2}(\vec{\sigma}\cdot\vec{\omega})\psi \quad (2.3)$$

___________
* Work supported by U.S.Department of Energy under the contract No. DE-AC02-76CH03000 and the National Science Foundation under Grant 1505458
[†] meiqin@fnal.gov

The rotation from $\theta_i$ to $\theta_f$ is expressed by a unitary matrix $M$ as

$$\psi_f = M\psi_i = e^{-i(\vec{n}\cdot\vec{\sigma})\phi/2}\psi_i \quad (2.4)$$

where $\vec{n}$ is the rotation axis, $\phi$ is the spin rotation angle. $\psi_i$ is the initial spin state, $\psi_f$ is the final spin state.

## SPIN TRACKING USING PTC

### Tracking Code PTC

The Polymorphic Tracking Code (PTC) [5] written by Étienne Forest is a library of Fortran90 data structures and subroutines for integrating the equations of orbital and spin motion for particles in modern accelerators and storage rings. PTC implements the high energy physics lattices and uses the "Fully Polymorphic Package", FPP, as the engine to do the Lie algebraic calculations. FPP implements Taylor maps (aka Truncated Power Series Algebra or TPSA) and Lie algebraic operations, which allows it to extract a Poincaré map from PTC. FPP also provides the tools to analyze the resulting map. The most common and most important tool is the **normal form**: with this at hand, one can compute tunes, lattice functions, and nonlinear extensions of these and all other standard quantities of accelerator theory. Indeed, the combination of PTC and FPP gives access to all of standard perturbation theory on complicated accelerator lattice designs.

### Normal form for spin on the closed orbit: $\vec{n}_0$

In PTC, spin is considered as a spectator, the closed orbit does not depend on spin. A map: $T = (m, S)$, where $m$ is an orbital map and $S$ is a spin matrix that depends on the orbit. This map acts on a ray $\vec{z}$ and a spin vector $\vec{s}$ as

$$T(\vec{z}, \vec{s}) = (m(\vec{z}), S(\vec{z})\vec{s}) \quad (3.1)$$

The matrix for the spin is evaluated at $\vec{z}$ and multiplies onto the vector $\vec{s}$. If a beam line #1 is followed by beam line #2, the spin map for the full beam line is then given by

$$T_2 \circ T_1 = (m_2, S_2) \circ (m_1, S_1)$$
$$= (m_2 \circ m_1, S_2 \circ m_1 S_1) \quad (3.2)$$

The matrix $S_2 \circ m_1 S_1$ is simply the product of $S_2 S_1$ where $S_2(\vec{z})$ is evaluated at $\vec{z} = m_1(\vec{z})$ with $\vec{z} = (x, p_x, y, p_y, z_5, z_6)$. If the map is a one-turn map around the closed orbit at some position s whose coordinates will be $\vec{0} = (0, 0, \cdots, 0)$, without loss of generality, it is straight forward to raise $T_s$ to a power $T_s^k(\vec{0}, \vec{s}) = (\vec{0}, S_s^k(\vec{0})\vec{s})$. This simply reflects the fact that on the closed orbit, the matrix $S$ for the spin is a constant matrix turn after turn. This matrix is a rotation and thus contains an invariant direction denoted as $\vec{n}_0$. We have $S_s^k(\vec{0})\vec{n}_0 = \vec{n}_0$. Now at some arbitrary position s, the matrix $S(\vec{0})$ can be expressed in terms of $\vec{n}_0$ and its rotation angle $\theta_0$ around $\vec{n}_0$:

$$S_s^k(\vec{0}) = \exp(k\theta_0 \vec{n}_0 \cdot \vec{L}) \quad (3.3)$$

The matrices $L_i$ are the usual generator of rotations obeying the commutation relations of the rotation group:

$$[L_i, L_j] = \varepsilon_{ijk} L_k \quad (3.4)$$

They are

$$L_1 = \begin{pmatrix} 0 & 0 & 0 \\ 0 & 0 & -1 \\ 0 & 1 & 0 \end{pmatrix} \quad L_2 = \begin{pmatrix} 0 & 0 & 1 \\ 0 & 0 & 0 \\ -1 & 0 & 0 \end{pmatrix} \quad L_3 = \begin{pmatrix} 0 & -1 & 0 \\ 1 & 0 & 0 \\ 0 & 0 & 0 \end{pmatrix} \quad (3.5)$$

where $\varepsilon = -1$ is chosen in the FPP package. $L_{1,2,3}$ are referred to as $L_{x,y,z}$ most of the time. Figure 3.1 gives a pictorial view of the algorithms of PTC. The red dot represents a ray moving in the "real world." The blue dots represent the spin.

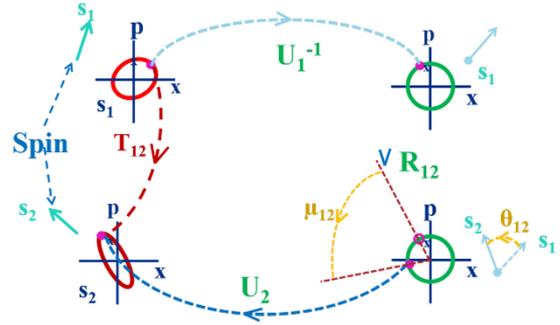

Fig. 3.1 Pictorial view of the algorithms of PTC.

### The nonlinear normal form for the invariant spin field (ISF): $\vec{n}(z)$

The invariant spin field (ISF) [6] was introduced by Barber and his collaborator as follows: there exists a vector $\vec{n}(\vec{z}, s)$, a 3–vector *field* of unit length obeying the T–BMT equation along particle orbits $(\vec{z}(s); s)$ and fulfilling the periodicity condition $\vec{n}(\vec{z}, s + C) = \vec{n}(\vec{z}, s)$, where $C$ is the circumference. Thus

$$\vec{n}(\vec{m}(\vec{z}, s); s + C) = \vec{n}(\vec{m}(\vec{z}, s); s)$$
$$= S_{3\times 3}(\vec{z}; s)\vec{n}(\vec{z}; s) \quad \text{or} \quad S\vec{n} = \vec{n} \circ \vec{m} \quad (3.6)$$

where $\vec{m}(\vec{z}; s)$ is the new phase space vector after one turn starting at $\vec{z}$ and s and $S_{3\times 3}(\vec{z}; s)$ is the corresponding spin transfer matrix. This equation states that a vector $\vec{n}(z)$ whose transformation is under the spin matrix $S(\vec{z})$ is the same as its transformation under the map $\vec{m}(\vec{z}; s)$. This equation can be easily applied to $\vec{n}_0$ since it is a constant under the application of $S(\vec{0})$ and the closed orbit is by the definition a constant, i.e. $\vec{m}(\vec{0}) = \vec{0}$. For an arbitrary $\vec{z}$, Eq.

(3.6) implies that if we follow $\vec{n}(\vec{z},s)$ after $k$ turns, the answer is simply $\vec{n} \circ m^k$. Thus the Fourier spectrum of $\vec{n} \circ m^k$ will not contain the spin frequency. This object behaves as if spin motion did not exist. If viewed as a vector field, the entire three dimensional field $\vec{n}(\vec{z},s)$ is left invariant under the action of the full spin-orbital map $T$. Obviously, if a particle at coordinate $\vec{z}=(x,p_x,y,p_y,z_5,z_6)$ starts with a spin slightly different from $\vec{n}(\vec{z},s)$, the actual spin will move around the axis $\vec{n}(\vec{z},s)$ and its spectrum will contain the spin tune as well as the orbital tunes. The chief aspects of the ISF are that:
1) For a turn–to–turn invariant particle distribution in phase space, a distribution of spins initially aligned along the ISF remains invariant (in equilibrium) from turn–to–turn, 2) For integrable orbital motion and away from orbital resonances, the ISF determines the maximum attainable time averaged polarization

$$P_{\lim} = \left|\langle \vec{n}(\vec{z},s)\rangle\right| \quad (3.7)$$

on a phase space torus at each $s$, where $\langle\ \rangle$ denotes the average over the orbital phases, 3) Under appropriate conditions, $J_s = \vec{n}\cdot\vec{S}$ is an adiabatic invariant while system parameters such as the reference energy are slowly varied.

## SPIN TRACK IN THE MAIN INJECTOR

The beam in the MI is injected at M306 (see Figure 1.1) from the Recycler Ring at an energy of 8 GeV, and accelerated to 120 GeV from 0.413 seconds to 1.08 seconds, then slow spilled for another 0.5 seconds to the extraction Septum at M522 to the fixed target experiment. The Siberian snake would be placed at M222, a straight section with a more than 10 m long drift space, opposite of the ring to M522. For the purpose of spin tracking in the MI, a special module called z_fnal_meiqin.f90 was written and added into the PTC library. It handles the acceleration of the proton beam through the γ-transition from a kinetic energy of 8 GeV at injection to the flat-top of 120 GeV (γ=9.528 to 128.93). Based on the MI ideal lattice, the transition energy γ is 21.619 at a time of 0.568 seconds after injection, as calculated by PTC. Furthermore, the real 21Cycle tables of the acceleration rate, the tunes and the chromaticity changes during the ramp, were also implemented in the module. There are 20 RF cavities in the MI for acceleration. They altogether are treated as one thin element at the end of the cavity section in PTC. The RF phase is 23.189° before the transition and (π-23.189°) right after the transition. The beam can be assigned by 95% of normalized emittance in the transverse planes and momentum deviations (Δp/p) in the longitudinal plane. Longitudinal emittance will then be calculated in the module.

With the help of Étienne Forest, a code in Fortran90, named fnal_injector_accelerate.f90, was written to do the orbit-spin tracking in the MI. We started with the flat output file of the latest MAD lattice of the Main Injector ring. The input file of the MAD lattice file was translated with Bmad [7] developed by David Sagan, Cornell University. Dave Sagan implemented the PTC/FPP library of Étienne Forest into his Bmad code. Therefore, they built up an interface between Bmad and PTC. MAD, Bmad and PTC agree to within machine precision.

After the orbit and spin were tracked for the first turn, the One-Turn-Map for both orbital and spin was obtained. Then the normal form for spin on the closed orbit: $\vec{n}_0$, was calculated, which actually is the Invariant Spin Field (ISF) on the closed orbit. Then, the spin polarizations of all particles at injection are chosen to be aligned with $\vec{n}_0$. A numerical computation of the ISF by stroboscopic average is compared with an evaluation of the normal form ISF of a single particle, and was found that there is very little difference between these two results.

### Tracking with an idea lattice

Multi-particle tracking was done first by sending 128 particles uniformly distributed (called "flat distribution" here) in vertical and longitudinal phase space, as seen in Figure 4.1. These 128 particles represent the beam. Using these results, the average polarization for different beam distributions can be calculated by integration as follows:

$$P_{ave} = \frac{1}{\sigma_1\sigma_2}\sum_{i=1}^{128}\rho(\varepsilon_1,\varepsilon_2)\vec{S}_i\Delta\varepsilon_1\Delta\varepsilon_2 \quad (4.1)$$

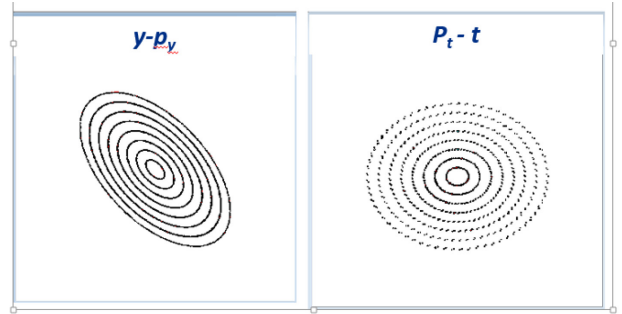

Fig. 4.1 Particles distributed in the phase space.

For a Gaussian distribution,

$$\rho(\varepsilon_1,\varepsilon_2) = \frac{1}{\varepsilon_1\varepsilon_2}e^{-\varepsilon_1/o_1}e^{-\varepsilon_2/o_2} \quad (4.2)$$

Then,

$$P_{ave} = \frac{1}{\sigma_1\sigma_2}\sum_{i=1}^{128}e^{-\varepsilon_1/o_1}e^{-\varepsilon_2/o_2}\vec{S}_i\Delta\varepsilon_1\Delta\varepsilon_2 \quad (4.3)$$

First, the particles were tracked with no snake. Figure 4.2 presents the results of the average polarization. The x-axis is $\gamma = E/E_o$, $E_o = 0.938 GeV$, and y-axis is the Polarization, "1" represents 100% polarization aligned with $\vec{n}_0$, the Invariant Spin Field on the closed orbit. The black line is the result of the initial beam with uniform distribution (Flat), and the red line represents the result of the initial beam with Gaussian distribution (cut at 6σ). Both cases show that polarization will be lost soon after the beginning of the acceleration.

The particles were then tracked with a single snake placed in the ring. The snake is represented by a point-like spin flipper. Figure 4.3 presents the average polarization of particles with 14 πmm·mrad (top panel) and 20 πmm·mrad (bottom panel) in vertical phase space, respectively. The momentum spread in both simulations is 1.25E-3. Polarization can be preserved at more than 90% at the end of the acceleration for most phase space distributions, except for the "Flat" distribution, which ends up at 88.8%. Similar results of average polarization were obtained for particles with 20 πmm·mrad in vertical phase space and momentum spread of 1.25E-3. The simulations show that the strongest resonance happens at γ=119.69 (Gγ=215), resulting in a big loss of polarization at the resonance, which however recovers in most cases. All the resonances here refer to the intrinsic spin resonances.

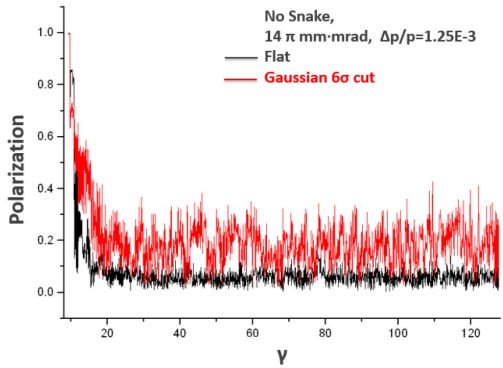

Fig. 4.2 Spin tracking with no snake in the ring for beams with different distributions in phase space.

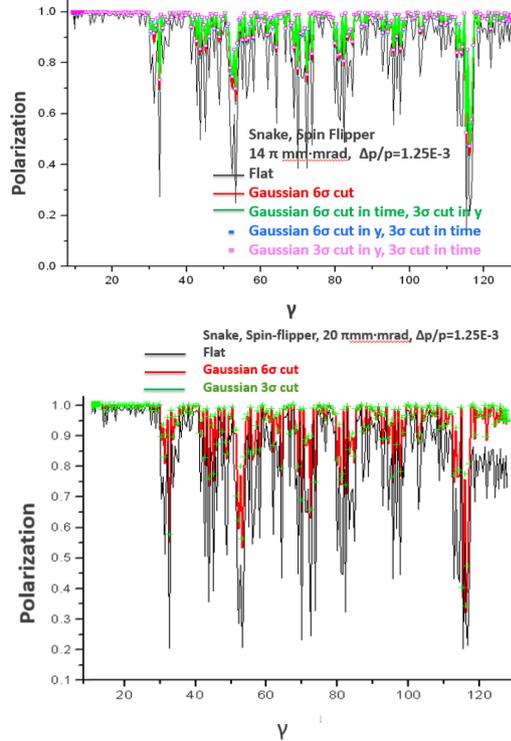

Fig. 4.3 Spin tracking in an ideal MI lattice, using a point-like snake to flip the spin. Top panel: the emittance of the beam is 14 πmm.mrad. Bottom panel: the emittance of the beam is 20 πmm.mrad. The distributions in phase space are flat, Gaussian with 3σ or 6σ cut in y (vertical plane) and in time (longitudinal plane).

*Tracking with a realistic lattice*

The measured field errors [8] and the misalignment data of all the magnets in the MI ring have been implemented into the MI lattice. PTC reads the survey coordinates and uses its pointer and patch functions to place each magnet into its actual position in the ring. This includes all the misalignment information, such as the shifts in x, y and z, as well as the roll and pitch angles.

The closed orbit of the MI at injection before correction due to the magnetic field errors and misalignment is shown in Figure 4.4

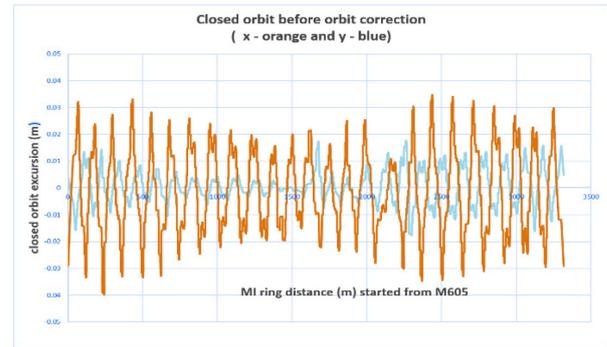

Fig. 4.4 Closed orbit before correction.

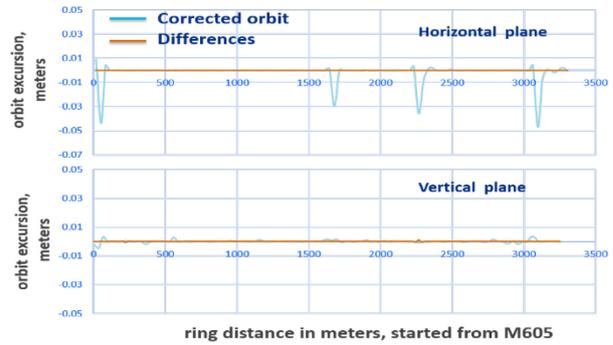

Fig. 4.5 Closed orbit after correction.

The orbit correction program for MI operation is able to correct the orbit to its desired orbit in RMS differences of 0.18 mm and 0.15 mm in the horizontal and vertical planes, respectively. PTC takes the beam position monitor (BPM) readings of the machine corrected orbit as the desired orbit, and corrects the closed orbit to it with almost no difference.

Spin tracking in PTC was performed after various orbit corrections have been applied. Just as in tracking with an idea lattice, PTC adjusts the tunes and chromaticities after the orbit corrections, and also during the ramp according to the ramp table of the MI operation 21 Cycler, the event for beams to SeaQuest experiment. 256 particles uniformly distributed on the vertical phase space with an emittance of 20 πmm·mrad and on the longitudinal phase space with a momentum spread $\Delta p/p = 1.25 \times 10^{-3}$ were tracked. Figure 4.6 presents the results. The *x-axis* is Gγ, with G=1.793 for protons, and γ the energy. The blue line represents the polarization for a beam with a flat particle distribution, while the orange line for a beam with a Gaussian distribution 6σ

Cut. The final average polarization remains at 85.2% for the flat distribution, and at 88.1% for the Gaussian distribution 6σ Cut. This presents the best results that the MI can achieve for closed orbit corrections.

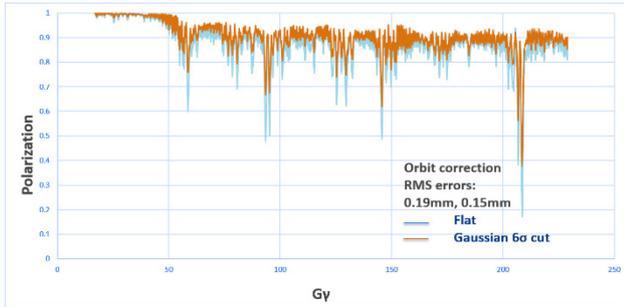

Fig. 4.6 Average polarization after PTC corrected the closed orbit to the BPM readings. The RMS values of the differences are almost zero. Note that the RMS values of the difference between the BPM readings and the MI desired positions are 0.19 mm in the horizontal and 0.15 mm in the vertical plane. Final polarization is 85.2% for the flat distribution and 88.1% for the Gaussian distribution 6σ cut.

The base tune of the MI is 26.425 in the horizontal plane and the 25.415 in the vertical plane. The spin depolarizing resonances occur when Gγ= mP±υ$_y$, where P is the periodicity of the lattice, and υ$_y$ is the vertical tune with υ$_y$=25.415. In the presence of magnetic field errors, periodicity is broken, P=1. Therefore, the most imperfect resonances occur for integers closest to ±υ$_y$. The tracking results show that the most imperfect resonances happen at Gγ=59, 94, 95, 146, 206, 209, which is near to M υ$_y$, , M=2, 4, 6, 8. The strongest resonance in this case is at Gγ= 209.

To see the effect of vertical orbit errors on the polarizations, we let PTC correct the closed orbit to the MI desired orbit with slightly larger RMS errors. Figure 4.8 presents the results for three different cases. In the top panel, the correction of the RMS error is 0.19 mm in the horizontal and 0.21 mm in the vertical planes. Not only do the depolarization resonances shown in Fig. 4.7 get stronger, but the two moderate resonances near Gγ=125 combined into one stronger resonance. Similarly, this also happened for resonance near Gγ=150 and 175. In the middle panel, the polarization is lost completely at Gγ=90 when PTC corrected the closed orbit to the MI desired orbit for RMS errors of 0.27 mm in the horizontal plane and 0.28 mm in the vertical plane. At Gγ=89.6, the average polarization drops below 70%. In the bottom panel, RMS errors of 0.43 mm in the horizontal plane and 0.45 mm in the vertical plane lead to strong imperfect resonance near Gγ=59 and 58.4, resulting in complete loss of polarization.

## CONCLUSIONS

Spin tracking of polarized protons in the Main Injector has been carried out for a "realistic" lattice that includes measured magnet field errors and misalignment survey data, as well as various degrees of orbit corrections that demonstrate the requirements that are needed to preserve significant polarization in the Main Injector. The simulations have shown that the polarization in the Main Injector is very sensitive to the excursions of the closed orbit. RMS deviations between the corrected orbit and the desired orbit should not be larger than 0.2 mm in both the horizontal and the vertical planes. The present MI orbit correction program can get closed orbit corrected to the desired orbit with RMS excursions of 0.18mm in the horizontal plane and 0.15mm in vertical plane. The final polarizations in the MI after ramping the energy up to 120 GeV can be kept at 85.2% for a flat beam distribution and 88.1% for a Gaussian beam distribution a 6σ cut using a point-like snake (flip-spin).

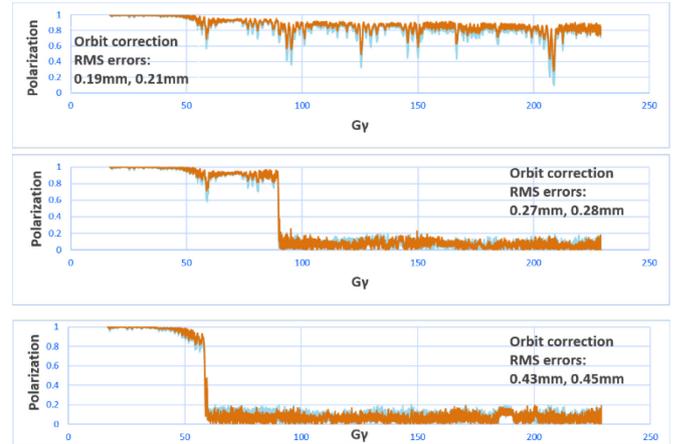

Figure 4.7 Average polarization for PTC closed orbit corrections to the desired orbit for various RMS values.

Implementation of a full-size 4-twist helical dipole instead of a point-like snake has started. PTC has successfully performed Symplectic integration through a specific 4-twist helical dipole configuration at various energies. Spin tracking studies of polarized protons in the Booster, as well as in the transfer lines are underway. This will allow to confirm whether polarized protons can be produced and maintained in the Fermilab accelerator complex using single Siberian snakes in the larger synchrotrons.

## ACKNOWLEDGEMENT

We would like to express appreciation to the Fermilab Accelerator Division, in particular to Sergei Nagaitsev and Ioanis Kourbanis, for their support. We would like to thank Étienne Forest (aka Patrice Nishikawa, KEK) for allowing us to use his remarkable PTC/FPP library and for his patience in explaining it to us in detail. We would also like to thank David Sagan (Cornell University) for use of his Bmad code to convert the MAD lattice format to a PTC readable format.